\documentclass{emulateapj}
\newcommand{\gsim}{\mbox{\raisebox{-1.0ex}{$~\stackrel{\textstyle >}
{\textstyle \sim}$ }}}
\newcommand{\be}{\begin{equation}}
\newcommand{\ee}{\end{equation}}
\newcommand{\ba}{\begin{eqnarray}}
\newcommand{\ea}{\end{eqnarray}}
\newcommand{\nn}{\nonumber}

\newcommand{\bt}{\begin{tabular}}
\newcommand{\et}{\end{tabular}}
\newcommand{\bc}{\begin{center}}
\newcommand{\ec}{\end{center}}
\newcommand{\ben}{\begin{enumerate}}
\newcommand{\een}{\end{enumerate}}
\newcommand{\bd}{\begin{description}}
\newcommand{\ed}{\end{description}}
\newcommand{\bit}{\begin{itemize}}
\newcommand{\eit}{\end{itemize}}
\newcommand{\ep}{\epsilon}

\newcommand{\epe}{\epsilon_{e}}
\newcommand{\epp}{\epsilon_{p}}
\newcommand{\epn}{\epsilon_{n}}

\newcommand{\nfe}{n_{f}^{e}}
\newcommand{\nfp}{n_{f}^{p}}

\newcommand{\mps}{m_{p}^{*}}
\newcommand{\mns}{m_{n}^{*}}
\newcommand{\mbps}{ \bar{m}_{p}^{*}}
\newcommand{\mbns}{ \bar{m}_{n}^{*}}
\newcommand{\gm}{\gamma}
\newcommand{\gme}{\gamma_{e}}
\newcommand{\gmp}{\gamma_{p}}
\newcommand{\gmn}{\gamma_{n}}

\newcommand{\kp}{\kappa_{p}}
\newcommand{\kn}{\kappa_{n}}
\newcommand{\sze}{s_{z}^{e}}
\newcommand{\szp}{s_{z}^{p}}
\newcommand{\szn}{s_{z}^{n}}

\newcommand{\met}{\tilde{m}_{e}}
\newcommand{\mpt}{\tilde{m}_{p}}
\newcommand{\mnt}{\tilde{m}_{n}}

\begin{document}

\shorttitle{Magnetic Domains in Magnetar Matter as an Engine for SGRs and AXPs}
\shortauthors{Suh \& Mathews}

\title{Magnetic Domains in Magnetar Matter as an Engine for
    Soft Gamma-ray Repeaters and Anomalous X-ray Pulsars}

\author{In-Saeng Suh\altaffilmark{1} and Grant J. Mathews\altaffilmark{2}}
\affil{Center for Research Computing$^{1}$ and Center for Astrophysics, Department of Physics$^{1,2}$, \\
University of Notre Dame, Notre Dame, IN 46556}

\altaffiltext{1}{Email: isuh@nd.edu}
\altaffiltext{2}{Email: gmathews@nd.edu}

\slugcomment{Accepted for publication in ApJ}

\begin{abstract}
Magnetars have been suggested as the most promising site for the origin of observed soft gamma-ray
repeaters (SGRs) and anomalous X-ray pulsars (AXPs).
In this work we investigate the possibility that SGRs and AXPs might be observational evidence
for a magnetic phase separation in magnetars. We study magnetic domain formation
as a new mechanism for SGRs and AXPs in which magnetar-matter separates into two phases
containing different flux densities.
We identify the parameter space in matter density and magnetic field strength at which there is an instability
for magnetic domain formation.
We conclude that such instabilities will likely occur in the deep outer crust for the magnetic
Baym, Pethick, and Sutherland (BPS) model and
in the inner crust and core for magnetars described in  relativistic Hartree theory.  Moreover,
we estimate that the energy released by the onset of this instability is comparable with
the energy emitted by SGRs.
\end{abstract}


\section{Introduction}

Soft gamma-ray repeaters (SGRs) are compact objects undergoing episodic instabilities
which produce  super-Eddington X-ray outbursts.
Up to now $6$ SGRs (4 confirmed and 2 candidates) have been observed\footnote{http://www.physics.mcgill.ca/$\sim$pulsar/magnetar/main.html}.
They are believed to be a new class of $\gamma$-ray transients
that are different from the source of ordinary gamma-ray bursts.
Observations of the spin-down time-scale \citep{kouveliotou1} have confirmed
the fact that these SGRs are newly born neutron stars with a very
large surface magnetic field ($B \sim 10^{15}$ G).
Such stars have been named magnetars \citep{duncan, kouveliotou2}.
About 10 AXPs are also categorized as magnetars \citep{kaspi, paradijs}.
Even though the magnetar model is generally accepted as the paradigm for both SGRs and AXPs,
it is not easy to explain both objects simultaneously by a consistent set of parameters.

Woods et al.\citet{woods} have reported evidence for a sudden magnetic field reconfiguration
in SGR 1900+14 during the giant flare of August 27, 1998. This scenario requires a reorganization
of the magnetic field both inside and outside the  star.
Sharp field gradients are postulated to  create a fracture of the rigid outer crust of the neutron star.
Cheng et al.\citet{cheng} have shown that SGR events and earthquakes share four distinctive statistical
properties: 1) power-law energy distributions; 2) log-symmetric waiting time distributions; 3)
strong positive correlations between waiting times of successive events; and 4) weak or no
correlation between intensities and waiting times. These statistical similarities, together
with the fact that the crustal energy liberated by starquakes is sufficient in principle
to fuel the soft gamma-ray flashes, suggest that SGRs are indeed powered by starquakes.
Moreover, there is also a  strong correlation between the magnitude and  waiting times
both for active earthquake regions and for SGR 1806-20. Thus, the statistical similarities
between earthquakes and SGR events argue for physically similar origins.
The Quasi-Periodic Oscillations (QPOs) observed at late-times of giant flares of SGR 1806-20, SGR 1900+14,
and SGR 0525-66 \citep{mareghetti, watts}
constitute  another piece of observational evidence in favor of  starquakes.
These QPOs are most likely due to seismic oscillations induced by the large crustal fractures occurring
in extremely energetic events similar to what happens after earthquakes.
Such oscillations could be limited to the crust or involve the entire neutron star.

Crackling noise \citet{sethna} arises when a system responds to changing external conditions through
discrete, impulse events spanning a broad range of sizes. Bak, Tang, and Wiesenfeld (1988) introduced
a connection between dynamical critical phenomena and crackling noise.  They emphasized how systems
may end up naturally at the critical point through a process of self-organized criticality.
Based upon this idea of the crackling noise, Kondratyev  (2002)
studied the statistics of magnetic noise
in neutron star crusts, and compared its intensity and statistical properties to the burst activity of SGRs.
He then argued that the noise could originate from magnetic avalanches.
However, because of the required inhomogeneous crust structure, he postulated the existence of magnetic domains within the neutron-star crust for  an interior magnetic field strength in the range $\sim 10^{16} - 10^{17}$ G. Using the randomly jumping interacting moment (RJIM) model, it was shown that the burst intensity and waiting time distributions are not only  in good agreement with observations, but also are analogous with the statistical properties of SGRs.

Whether or not  magnetars are the source of SGRs and AXPs,
as relics of stellar interiors, the study of the magnetic fields in and around
degenerate stars should give important information on the role such fields play in
star formation and stellar evolution \citep{suh01}.
The scalar virial theorem implies an allowed internal field strength for a star of
$B \lesssim 2 \times 10^8 (M/M_{\odot}) (R/R_{\odot})^{-2}$ G
for a star of size $R$ and mass $M$. For a typical neutron star
the maximum interior field strength could thus reach $B \sim 10^{18}$ G.
Since strong interior magnetic fields modify the nuclear equation of state for degenerate
stars, their structure will also be changed \citep{cardall}.

Even though SGRs appear to be observable consequences of starquakes or surface fractures,
their detailed mechanism is not known. Moreover, it is unlikely that AXPs could only be explained by
starquakes in strong magnetic fields. Therefore, in this paper we suggest a magnetic domain model  to correlate smoothly between the statistics of starquakes and magnetic avalanches in magnetar crusts.
In this work magnetic properties of magnetar-matter such as the magnetization
and the susceptibility are calculated in the framework of three different representative equations of state.
We consider an ideal $n-p-e$ gas, relativistic Hartree mean field theory, and
the magnetic Baym, Pethick, and Sutherland (BPS) model \citep{lai}.
 It has been shown \citep{broderick} that the magnetization of magnetar-matter undergoes large $de ~Haas-van ~Alphen$ oscillations.
The magnetic susceptibility can then lead to a  region unstable to the formation of  {\it magnetic domains}.
It has not yet been demonstrated, however,  that magnetic domains actually form in magnetar-matter.
Here we show that it is indeed possible to form such magnetic domains  in magnetars, and that  these could affect the surface properties and structure of magnetars  possibly leading to observable consequences such as starquakes, glitches,  and $X$- or $\gm$-ray emission.

\section{The Differential Susceptibilities for a Magnetic $npe$ gas}
The magnetic equation of state for magnetar-matter has been described in Suh \& Mathews (2001a)
for an ideal $npe$ gas.
The magnetization of simple  magnetar-matter material can be derived from the thermodynamic potential
\citep{blandford}.   Brodrick et. al. (2000) have generalized the formalism  for a multicomponent system
including interacting nucleons. Hereafter we introduce the following notation for magnetic fields.
For material in a uniform magnetic field, the magnetic field $H$ is related
to the flux density $B$ by the relation \citep{pippard}:
\be
B = H + 4 \pi (1 - {\cal D}) {\cal M}(B)~~,
\ee
where ${\cal D}$ is the demagnetization coefficient that is fixed by the
geometry of the system. For example, ${\cal D} \approx 0$ for a neutron star crust permeated
by an approximately vertical magnetic field. In chemical equilibrium, the total magnetization $\cal M$
is given by a simple sum of the constituent magnetizations.
\be
{\cal M} = \sum_{j= n,p,e}  {\cal M}_{j}~~.
\ee
The magnetic susceptibilities are then given by $\chi = {\cal M}/H$, and
the differential susceptibility $\eta$ is defined \citep{blandford} by
\be
\eta_j = \Bigg( \frac{\partial {\cal M}_j}{\partial B} \Bigg)_{\mu,T,V} ,
\ee
where $\mu$ is the chemical potential, $T$ is the temperature, and $V$ is the volume of the system.
The total differential susceptibility for magnetar-matter above the neutron drip density is then given by
\be
\eta = \sum_j \eta_j, \;\; where \;\; j=n,p,e~~.
\ee

For a cold ideal $npe$ gas, we can obtain simple expressions for the differential susceptibilities of
the various components.
For the electron differential susceptibility $\eta_{e}$ we have,
\ba
\eta_{e} &=&  \frac{\alpha}{4 \pi^2}
\sum_{\sze, \, n} \Bigg[ 4 \nfe {\rm ln} \Bigg(\frac{\epe + \sqrt{\epe^2 - \met^2}}{\met} \Bigg) \nn \\
&~& - 2 \, \gme \Bigg(\frac{\nfe}{\met}\Bigg)^2  \Bigg( \frac{\epe}{\sqrt{\epe^2 - \met^2}} \Bigg)
\Bigg]~~,
\ea
where $\met = \sqrt{1 + 2 \gme \nfe}$.
For protons,
\ba
\eta_{p} &=& \frac{\alpha}{4 \pi^2}
\sum_{\szp, \, n} \Bigg[ \Bigg\{ 4 {\cal A}_{p}
- 2 \gmp \frac{\nfp{^2}}{(1 + 2 \gmp \nfp)^{3/2}} \Bigg\} \nn \\
  &+& \, \gmp {\cal A}_{p}^2
\Bigg\{ {\rm ln} \Bigg(\frac{\epp + \sqrt{\epp^2 - \mpt^2}}{\mpt} \Bigg)
- \frac{\epp}{\sqrt{\epp^2 - \mpt^2}} \Bigg\} \Bigg],
\ea
where ${\cal A}_{p} = {\nfp}/{\sqrt{1 + 2 \gmp \nfp}} - \szp \kp$ and $\mpt = \sqrt{1 + 2 \gmp \nfp}$.
In Eqs. (5)-(6), $n_{f}^{e, \, p} = n + \frac{1}{2} - s_{z}^{e, \, p}, $ where $n = 1, 2, 3, ...$ denotes the Landau levels and
$s_{z}^{e, p}$ are the electron ($e$) and proton ($p$) spin projection on the magnetic field direction,
$\gamma_{e, p} = B/B_{c}^{e, \, p}$ where $B_c^{e, \, p} = e \hbar / m_{e, \, p}^{2} c^{3}$ are the quantum critical field for
electrons and protons, and $\ep_{e, \, p} = E_{F}^{e, \, p} / m_{e, \, p} c^{2}$ with the electron and proton Fermi energy
$E_{F}^{e, \, p}$, respectively.

Finally, for neutrons we obtain,
\be
\eta_{n} = \frac{\alpha}{2 \pi^2}
\sum_{\szn} (\szn \kn)^2 \Bigg[ \eta_{n}^{0} + \szn \kn \gmn \eta_{n}^{\kappa}\Bigg],
\ee
where
\be
\eta_{n}^{0} = - \frac{1}{2} \Bigg\{\epn\sqrt{\epn^2 - \mnt^2}
+ \mnt^2 {\rm ln}\Bigg(\frac{\epn + \sqrt{\epn^2 - \mnt^2}}{\mnt} \Bigg) \Bigg\},
\ee
and
\be
\eta_{n}^{\kappa} = \mnt {\rm ln} \Bigg(\frac{\epn + \sqrt{\epn^2 - \mnt^2}}{\mnt} \Bigg)
\ee
with $\mnt = 1 + 2 \szn \kn \gmn$ and $\gamma_n = B / B_{c}^{n}$,  $B_{c}^{n} = e \hbar / m_{n}^{2} c^{3}$.
In Eqs. (5)-(7), $\alpha = e^2 / \hbar c$ is the fine structure constant,
$s_{z}^{n}$ is the neutron spin projection in the magnetic field direction, and
$\kp, \, \kn$ are the anomalous magnetic moments for protons and neutrons respectively, as given below in Eq. (10).
Here we use the same notation as  in Suh \& Mathews (2001a) for the particle Fermi energy and magnetic field strength.

\section{The Differential Susceptibilities in the Relativistic Hartree Theory}

For a system of strongly interacting baryons (neutrons and protons), the relativistic mean field (Hartree)
theory should be a reasonable approximation for the description of the equation of state for magnetar-matter
at high density \citep{broderick,chakra} through the exchange of $\sigma$ and vector $\omega, \rho$ mesons
in a strong magnetic field. In the baryon Lagrangian for the relativistic Hartree theory,
the anomalous magnetic moments are included through the coupling
of the baryons to the electromagnetic field tensor with
$\sigma_{\mu \nu} = \frac{i}{2} \left[\gamma_{\mu},\gamma_{\nu} \right]$
and the strengths $\kp$ and $\kn$ given by
\be
\kappa_p = \frac{e}{2 m_{p} c} \Bigg(\frac{g_p}{2} - 1 \Bigg), \;\;\;\;\;
\kappa_n = \frac{e}{2 m_{n} c} \frac{g_n}{2}~~,
\ee
where $g_p = 5.58$ and $g_n = -3.82$ are the Lande $g$-factors for protons and
neutrons, respectively.
In this work, we can ignore the possible scalar $\sigma$, the vector $\omega$
and the iso-vector $\rho$ meson self-interactions. Therefore,
although the electromagnetic field is included in the total Lagrangian,
it assumed to be externally generated (and thus has no associated field equation)
and only frozen-field configurations will be considered.
The effective baryon mass $m_{b=p,n}$ is then given by the coupling to the $\sigma$ meson,
\be
m_b^{*} = m_b - (g_{\sigma}/{m_{\sigma}})^{2} (n_{p}^{S} + n_{n}^{S})~~,
\ee
where $g_{\sigma}$ and $m_{\sigma}$ are the $\sigma$ meson coupling constant and mass respectively.
In Eq. (11),
$n_{p}^{S}$ is the scalar number density for protons,
\ba
n_{p}^{S} &=& \frac{1}{2 \pi^2} \sum_{\szp, \, n}  \gmp^{*}
\Bigg( \frac{\mps c}{\hbar} \Bigg)^3 \nn \\
&\times& \frac{\mbps}{\mbps - \szp \kp \gmp^{*}}
{\rm ln} \Bigg(\frac{\epp^{*} + \sqrt{\epp^{* 2} - \mbps{^2}}}{\mbps} \Bigg)~,
\ea
where $\mbps = \sqrt{1 + 2 \nfp \gmp{^*}} - \szp \kp \gmp^{*}$ and $\gmp^{*} = (m_e / \mps)^2 \gme$, while
the scalar number density for neutrons is
\ba
n_{n}^{S} &=& \frac{1}{4\pi^2} \sum_{\szn} \Bigg( \frac{\mns c}{\hbar} \Bigg)^3
\Bigg[\epn^{*} \sqrt{\epn^{* 2} - \mbns} \nn \\
&-& \, \mbns{^2} \, {\rm ln} \Bigg( \frac{\epn^{*} + \sqrt{\epn^{*2} - \mbns{^2}}}{\mbns} \Bigg) \Bigg]~,
\ea
with $\mbns = 1 + \szn \kn \gmn{^*}$ and $\gmn^{*} = (m_e / \mns)^2 \gme$.
For simplicity, the nucleon rest mass is taken as $m = m_n = m_p$ in the numerical calculation \citep{chakra}.

Assuming a mixture of neutrons, protons, and electrons in chemical
equilibrium, the chemical potentials are related by
\be
\mu_n = \mu_p + \mu_e~~,
\ee
while the condition of charge neutrality gives
\be
n_p (\epp, \gmp) = n_e (\epe, \gme) \, .
\ee
Given the nucleon-meson coupling constant and the coefficients in the scalar self-interactions,
the field equations can be solved self-consistently for the chemical potentials, $\mu_{j} \; (j=n,p,e)$,
and the meson field strengths in a uniform magnetic field $B$
along the $z$ axis corresponding to the choice of the gauge for the vector potential $A^{\mu}$ \citep{broderick}.
In this work, we adopt the following coupling constants and mesons masses:
$g_{\sigma}^{2} (m_{N}/m{\sigma})^2 = 357.47$, $g_{\omega}^{2} (m_{N}/m{\omega})^2 = 273.78$,
and $g_{\rho}^{2} (m_{N}/m{\rho})^2 = 97.0$ \citep{horowitz}.

Figure 1 shows the effective baryon mass $m^{*} / m$ as a function of baryon density $\rho$
for magnetic field strengths of $\gme = 0.01$ (solid line) and $10^5$ (dash line) calculated in the model of Horowitz and Serot (1981).
For a magnetic field strength less than $\sim 10^{18}$ G, This figure shows that the effective nucleon mass
is not significantly affected by magnetic field strength.
Broderick et al. (2000) and Chakrabarty et al. (1997) have obtained similar results.
This effective baryon mass modifies the baryon dispersion relation in dense magnetar-matter.

\begin{figure}
\vspace{2.4cm}
\plotone{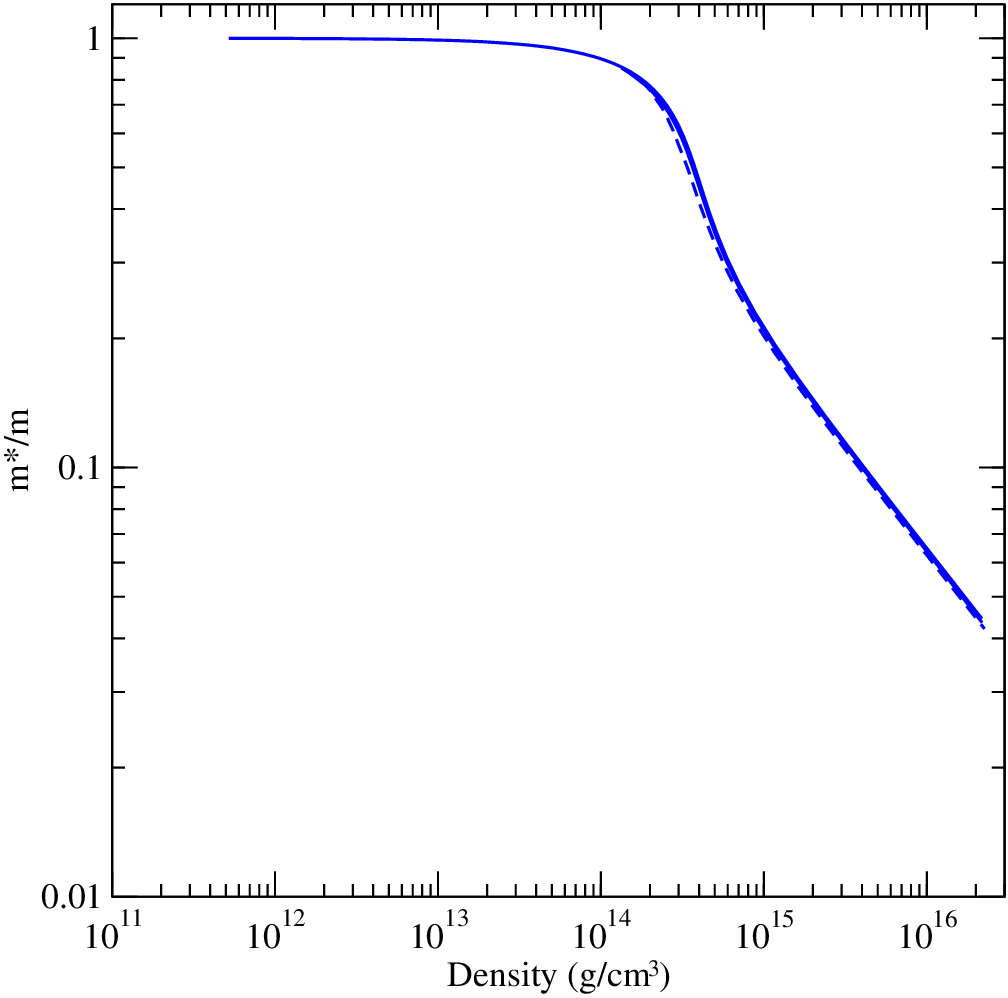}
\caption{Variation of the effective baryon mass $m^*/m$ in the relativistic Hartree model as a function of baryon density for magnetic
field strengths of $\gme = 0.01$ (solid line) and $10^5$ (dash line).
\\
}
\label{fig:rho-mstar}
\end{figure}

\section{The Differential Susceptibility in the Magnetic BPS Model}

For  matter in thermodynamical equilibrium below the neutron drip density, $\rho_{drip} \approx 4.3 \times 10^{11}$ g/cm$^3$,
we adopt the magnetic Baym, Pethick, and Sutherland (BPS) model \citep{lai}
and use the semi-empirical mass formula \citep{shapiro}. For simplicity, we only consider $^{56} _{26}$Fe nuclei
in the numerical calculation.
Then, the magnetization and the differential susceptibility for the
magnetic BPS equation of state are given by
\be
{\cal M}_{BPS} = {\cal M}_e + \frac{1}{3} {\cal M}_L , \;\;\;\;\;
\eta_{BPS} = \eta_e + \frac{1}{3} \eta_{L}
\ee
where ${\cal M}_e$ is the magnetization of the electron gas and $\eta_e$ is given in Eq. (5). In Eq. (16),
${\cal M}_L$ is the magnetization for the $bcc$ Coulomb lattice energy.
Then we drive here the lattice differential susceptibility $\eta_L$ to be,
\ba
\eta_L &=& - \; 1.444 \Bigg(\frac{1}{2 \pi^2} \Bigg)^{4/3} Z^{2/3} \alpha^{2} \;
\sum_{\sze, \, n} \frac{1}{(\epe^2 - \met^2)^{5/3} \gme^{2/3}} \nn \\
&\times& \Bigg[ (\epe^2 - \met^2)^{7/3} - 8 \gme \nfe (\epe^2 - \met^2)^{4/3} - 2 (\gme \nfe)^2 \Bigg]
\nn \\
\ea
where $Z$ is the average atomic number of  the nuclei.

\section{Magnetic Domain Formation}

\begin{figure}
\vspace{2.4cm}
\plotone{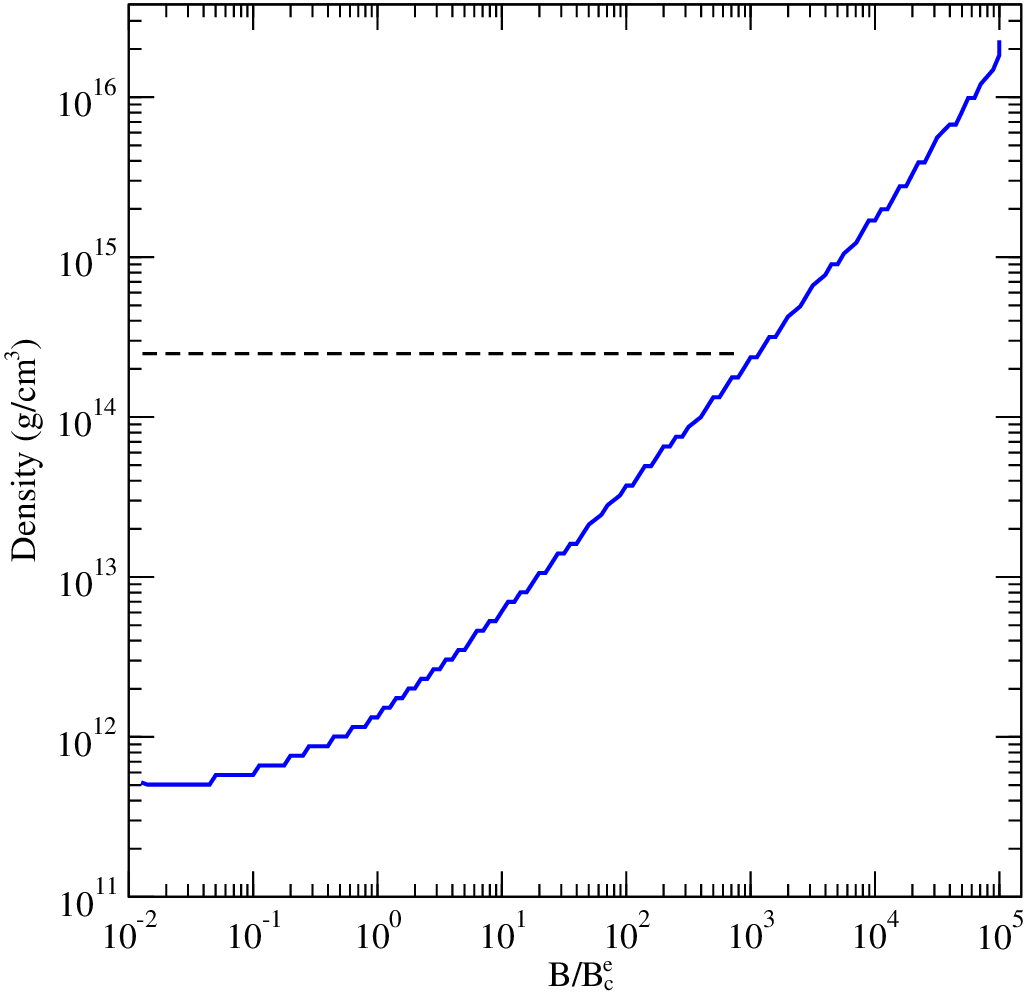}
\caption{Region of the magnetic domain formation instability for an ideal $npe$ gas as a function of
baryon density for various magnetic fields in units of $B_c^e = 4.414 \times 10^{14}$ G.
The region above the solid line denotes the parameter values for which conditions are unstable
for the formation of magnetic domains.
The dash line indicates the nuclear saturation density $\rho_{nuc} \approx 2.8 \times 10^{14}$ g/cm$^3$.
\\
}
\label{fig:B-rho-npe}
\end{figure}

In general, the magnetization of a system is small compared with the external magnetic field $H$.
However, when the system is sufficiently cool so that its thermal energy is smaller than the spacing
of the Landau levels, the magnetization can undergo large $de~ Haas-van ~Alphen$ oscillations with
either changing magnetic fields or a changing Fermi energy.
Under these conditions, it sometimes becomes energetically favorable for the system to separate
into two phases containing different flux densities. This  is the so-called Schoenberg effect \citep{pippard}.
This means that although $4 \pi \chi$ is less than unity for a magnetized $npe$ gas,
in certain regions $4 \pi \eta$ can exceed unity which implies the possible existence of
magnetic domains \citep{blandford}.
That is, when the differential susceptibility obeys $\eta > 1/4\pi$, and
$\partial H / \partial B < 0$.
Then  magnetar matter in thermodynamic equilibrium becomes unstable to the formation of magnetic
domains of alternating magnetization.
For the case of a vanishing demagnification coefficient, ${\cal D} = 0$, the material will separate into two
phases corresponding to different magnetization.

For magnetar-matter above the neutron drip density $\rho_{drip}$, we considered an ideal pure non-interacting cold $npe$ gas
as well as the relativistic Hartree model. However, in the density region between
$\rho_{drip}$ and $\rho_{nuc} \approx 2.8 \times 10^{14}$ g/cm$^3$, neutron-star matter is composed electrons, nuclei,
and free neutron gas so that we can not directly apply the ideal $npe$ gas model in this density regime.
For example, for non-magnetic neutron-star matter in this intermediate density regime,
we can employ the Baym, Bethe, and Pethick (BBP) equation of state \citep{bbp}.
This BBP model is based upon a compressed liquid drip model of nuclei. It gives some corrections
to the ideal $npe$ equation of state. Therefore, if we adopt the magnetic BBP model, the region below the dash line in
Figs. 2 and 3 will be shifted to the left because of the BBP equation of state [See \citet{shapiro}].
This means that for a fixed magnetic field strength the density region in which magnetic domains can be formed increases.
However, since there is no physical model in the intermediate density regime with a strong magnetic field,
we can simply describe this regime using an analogy from our ideal $npe$ gas model.

\begin{figure}
\vspace{2.4cm}
\plotone{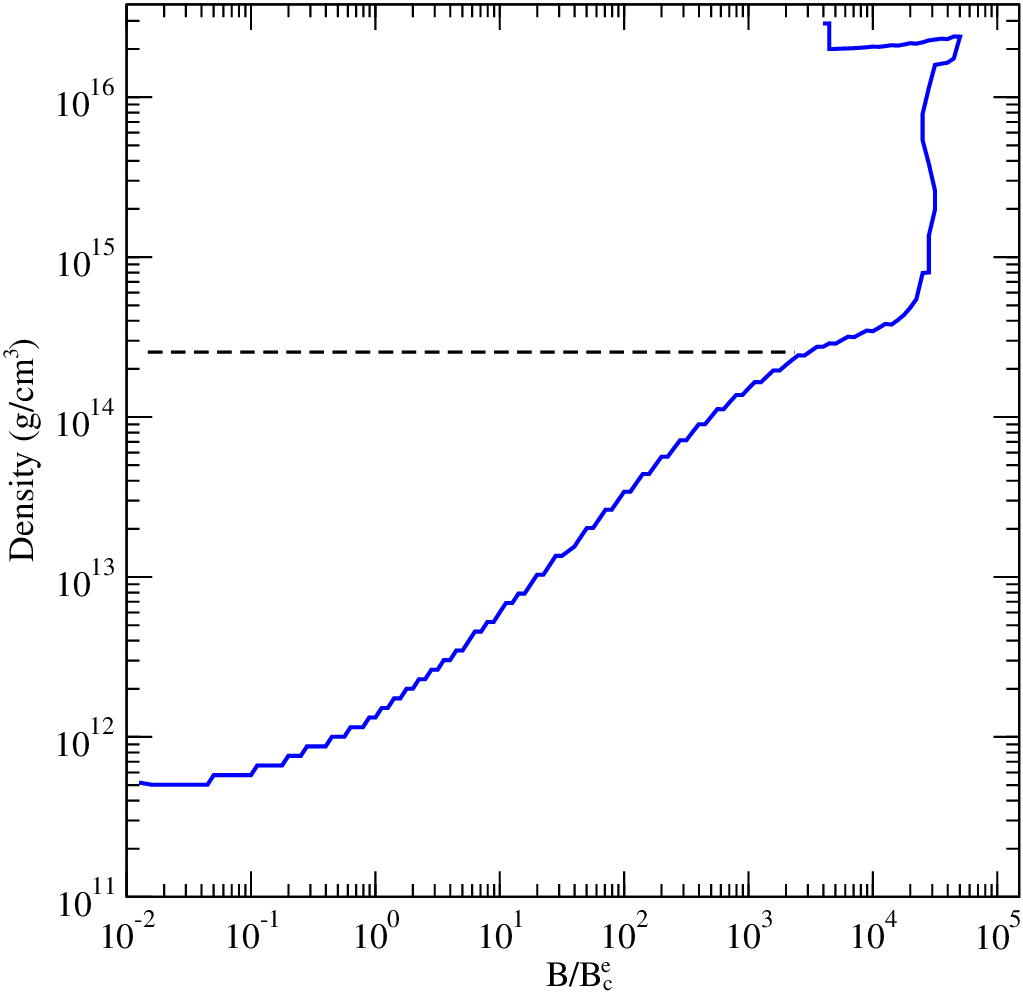}
\caption{Region of the magnetic domain formation instability for the relativistic Hartree theory
in which the effective baryon mass is taken into account.
The region above the solid line denotes the parameters for which the conditions are unstable
for the formation of magnetic domains.
The dash line indicates the nuclear saturation density $\rho_{nuc} \approx 2.8 \times 10^{14}$ g/cm$^3$.
\\
}
\label{fig: B-rho-mstar }
\end{figure}

Figure  2 shows the  regions  of  matter density and magnetic flux density for an ideal $npe$ gas above the $\rho_{drip}$
in which magnetar-matter is unstable to a phase separation into magnetic  domains.
For magnetic field strength of $\sim 10^{14} - 10^{15}$G,  magnetic domains cannot be formed in the lower density region
below $\sim 10^{13}$ g/cm$^3$.
However, it is possible for magnetic domains to form in the density region higher than $\sim 10^{13}$ g/cm$^3$.

Figure 3 shows the unstable region for a relativistic Hartree mean field model in which the baryon effective mass
is taken into account.
When we consider the effective baryon mass within the relativistic Hartree theory, domain formation can
significantly occur  above a density  of $\rho \sim 10^{14}$ g/cm$^{3}$ and a field strength $B \sim 10^{16}$ G.
The effective baryon mass lowers the density at which magnetic domain formation occurs in the
core of a magnetar in which strong magnetic fields are expected.
We also find that magnetic domain formation could not be formed for a  magnetic field strength of $\lesssim 10^{18}$ G.

\begin{figure}
\vspace{2.4cm}
\plotone{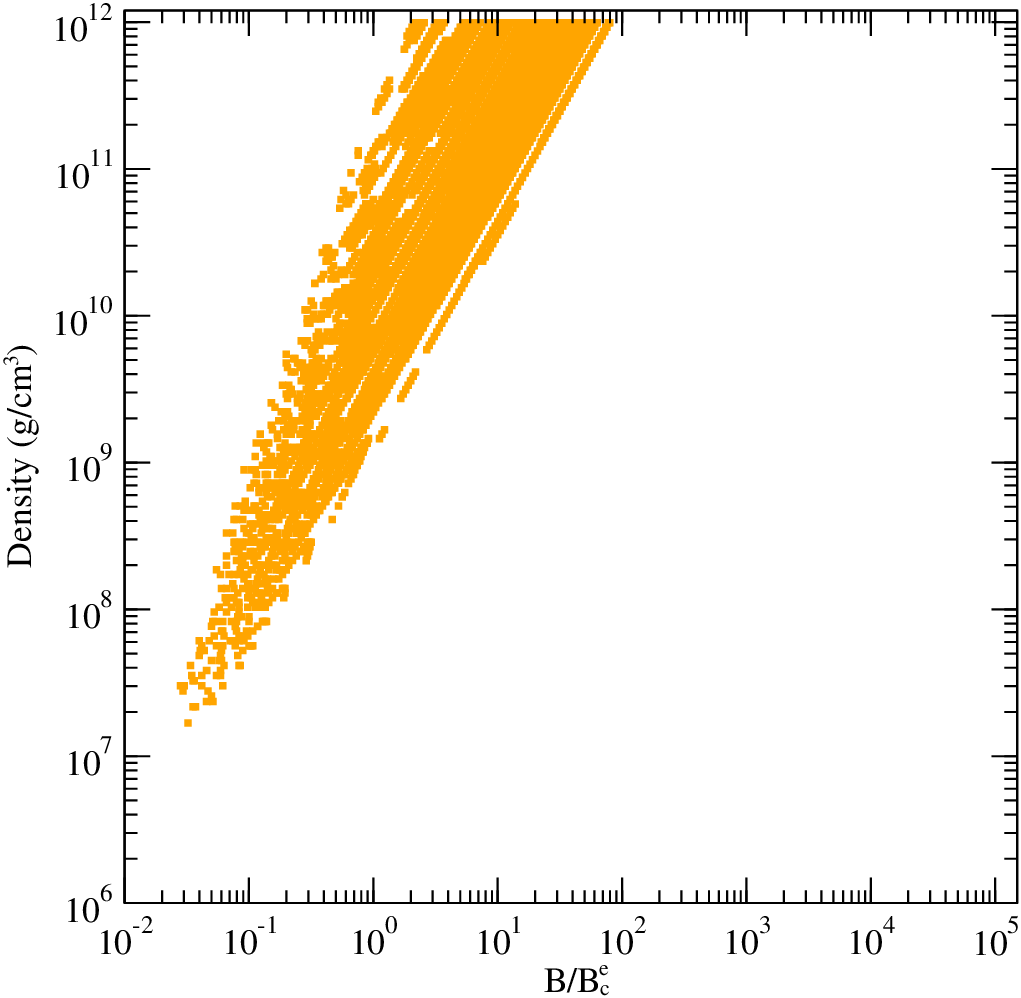}
\caption{Magnetic domain formation instability for $^{56}_{26}$Fe in the magnetic BPS model.
The shaded regions denote the parameter regions for which  conditions are unstable for the formation of magnetic domains.
\\
}
\label{fig: B-rho-BPS }
\end{figure}

Finally, in Figure 4,  the magnetic domain instability regions obtained by using the magnetic
BPS model are depicted in the outer crust of magnetars.  Unlike the other two cases, in this case the magnetization
is dominated by electrons.  Therefore the conditions for the onset of the magnetic domain instability is affected by
the occupation of Landau levels.  This causes the instability conditions to vary as a function of field strength for fixed density.
In Figure 4 we see that  magnetic domains cannot be formed if the density is less than
$\rho \sim 5 \times 10^{10}$ g/cm${^3}$ for a magnetar having a typical  surface magnetic field of $B \sim 10^{14 - 15}$ G.
However, in the density region around the neutron drip density, magnetic domains can be formed.
We can also  see magnetic domain formation in regions
at low density  $\gsim 10^{11}$ g/cm$^3$ and at a relatively low magnetic field strength
$\gsim 10^{13}$ G. This is the case considered by Blandford and Hernquist (1982).
For magnetic white dwarfs, Adam (1986) also obtained the similar result for a magnetized electron gas.

We thus find that possible unstable regions for magnetic domain formation are
within the deep outer crust, deeper part of the inner crust, and core
of magnetars. This means that the outer shell at low density less than $\rho \sim 10^{10}$ g/cm${^3}$ is stable
against magnetic domain formation and should consists of strongly magnetized material without magnetic domains.

\section{SGR and AXP Mechanisms}

This magnetic domain formation might be an important clue to explain SGRs and AXPs in the magnetar model.
As the density increases, the number of Landau orbital also increases. Therefore,
Based upon Blandford $\&$ Hermquist (1982), we can calculate the thickness of the crust, $\Delta z$,
using the gravitational potential energy of a nucleon and the Fermi energy $\epe$ per electron.
We also know the maximum number of the Landau orbitals $n  =  (\epe^2 - 1) / 2 \gme$,
where $\gme = B / B_{c}^{e}$. For a fixed B, $\Delta n \sim \epe \Delta \epe / B$.
Therefore, we can get $\Delta z / \Delta n$.
For $\Delta n = 1$, we finally can estimate the spacing between layers associated with
the maximum Landau level:
\be
\Delta z \sim 100 \Bigg(\frac{g}{10^{14} {\rm cm/sec^2}} \Bigg)^{-1}
\Bigg( \frac{\mu_e}{2} \Bigg)^{-1} \Bigg( \frac{\epe}{5} \Bigg)^{-1}
\Bigg(\frac{B}{10^{15} {\rm G}} \Bigg) \; {\rm m}~~,
\ee
where $g$ is the surface gravity, $\mu_e$ is the mean molecular weight per electron, and
$\epe = E_{F}^{e} /m_e c^2$ denotes the electron Fermi energy.
Under the conditions for domain formation, the spacing $\Delta z$ would be the average
vertical scale between domain interfaces as well as a horizontal scale of the domains
if they are in  local pressure equilibrium.
[However, the actual size and shape of the domains are difficult to determine.]
Now, for a fixed the Fermi energy, we get an equation $\Delta n$ which is given by $\Delta B$.
For $\Delta n = 1$ then the horizontal variation in the magnetic field will be
\be
\frac{\Delta B}{B} \sim 50 \Bigg(\frac{B}{10^{15} {\rm G}} \Bigg) \frac{1}{\epe^2 - 1} .
\ee
This implies that there are different magnetic flux densities in region
where magnetic domains can form.
However, in the outer crust at lower density, the magnetic fields are homogeneously
distributed and tightly pinned to the matter.

According to the magnetic domain theory \citep{pippard},
the magnetic domain walls move or grow, and the magnetic domains rotate within the material.
Therefore, the movement and rotation of magnetic domains can cause a physical dimensional change
and produces the maximum possible strain on the crust material.
This process is very similar to that of the internal structure of the earth
in which a sudden collapse or strain of the mantle below the earth's crust sometimes occurs.

In regions where magnetic field distortion increases, magnetic domains could be formed.
At the boundary there will also exist regions around each wall where the magnetic field is distorted.
The formation or adjustment of domain structure involves
magnetic field fluctuations of a few percent amplitude which have an
anisotropic magnetostrictive stress $2 \pi {\cal M}^2$ associated with the magnetization.
Any sudden readjustment of the domain structure will cause a local departure
from isostasy which will be relieved on an ohmic dissipation timescale
($\tau_D \sim \sigma A / 4 \pi c^2 \sim 10^{4}$ yr) \citep{blandford}.
These anisotropic magnetostrictive stresses may be large enough to crack the outer crust \citep{blaes}.

Then, we can estimate the physical length variation of the magnetic domains. The bulk modulus $(K_S)$ is defined
as the pressure increase needed to effect a given relative decrease in volume.
For a gas, the adiabatic bulk modulus $K_S$ is approximately given by $K_S = \Gamma P$, where $\Gamma$ is the
adiabatic index. Then,
\be
\frac{\delta V}{V} \approx \frac{\lambda}{K_S} {\cal M} H \approx 0.1 \lambda~~,
\ee
for $\Gamma \simeq 1.8$ and $\chi \simeq 1$ at the neutron drip density ($4.4 \times 10^{11}$ g/cm$^3$
in the magnetic BPS model \citep{lai}). Therefore, with $\lambda \approx 0.6$ for Fe \citep{stewart},
the length change due to magnetic domain formation is finally given by
\be
\frac{\delta l}{l} = \frac{1}{3} \frac{\delta V}{V} \approx 0.02~~.
\ee
This means there is a  2 $m$ length change for a 100 $m$ characteristic domain size when a magnetic domain is formed
in the deep outer crust of a magnetar.
Now we can estimate a cracking timescale in the outer crust to be
$\tau_C \sim \Delta z / v_s \simeq 0.1$ ms, where $v_{s} = \sqrt{Y/\rho}$ (with Y = shear modulus)
is the shear velocity \citep{blaes}. Kontratyev (2002) has analogized this cracking timescale as the avalanche spanning
time which is consistent with the rise time for SGR giant bursts \citep{hurley, kouveliotou2, mareghetti}.

Finally, we can estimate the elastic energy released, $\Delta E_{D} \sim \Delta z \, R_D^{2} \sigma$,
where $R_D \sim \Delta z$ is the characteristic horizontal size of the domain,
using the magnetic stress energy $\sigma \sim \, \chi^2 B^2$.
Then, we obtain the released elastic energy
\be
\Delta E_D
\sim 6 \times 10^{42} \chi^2 \theta_m \Bigg( \frac{R_D}{10^4 {\rm cm}} \Bigg)^2
\Bigg( \frac{B}{10^{15} {\rm G}} \Bigg)^2 {\rm erg}.
\ee
where $\theta_m$ the maximum allowed strain angle, and $\chi = {\cal M}/H$ is the magnetic susceptibility.
This is also the typical energy released in SGRs.
Hence, this cracking of the crust by magnetostrictive stress would be the
mechanism of the observed SGRs. We suggest that magnetic domain formation and any sudden readjustment
of the domains can produce an energy source for soft gamma-rays in  SGRs and X-rays in AXPs.

However, there is evidence that AXPs have stronger magnetic fields than SGRs \citep{mareghetti}.
With stronger magnetic fields, it would be hard for the magnetic domains to be formed in the outer crust of magnetar.
This means that there would be little  cracking of the outer crust by the magnetostrictive stress.
The possibility remains, however,  that  AXPs could produce a giant blast like SGRs in this magnetic domain model.

\section{Summary}

In this work we have studied magnetic domain formation as a new mechanism for SGRs and AXPs.
In this paradigm magnetar-matter separates into two phases containing different flux densities.
We have identified the parameter space in matter density and magnetic field strength at which there is an instability
for magnetic domain formation and have shown  that such instabilities are  likely to occur in the deep outer crust
for the magnetic BPS model, and in the deeper part of the inner crust and core for magnetars described by  relativistic Hartree theory.
Moreover, we have estimated the strain on the outer crust induced by the formation of such domains and found that
the anticipated energy release  is comparable with the energy emitted by typical SGRs.
Hence, we propose that the magnetic domain formation scenario described here represents a new possible mechanism to drive
the giant flares of SGRs as well as  X-ray outbursts and the quiescent phase of AXPs.
At the very least,  this proposal warrants further investigation.
Moreover, since the physical length variation caused by the magnetic domain formation might lead to solid crustal
deformation and catastropic cracking, SGRs might be sources of gravitational waves (GWs) \citep{abbott}
even though there is not yet evidence of GWs associated with  observed SGR bursts.
However, if it becomes possible to detect GWs from SGRs,
that may be a way to verify the magnetic domain model in magnetars.
Clearly, the next step is to undertake detailed dynamical numerical studies of the formation and evolution of such magnetic
domains in neutron star crusts.  Efforts along this line are currently underway \citep{suh03}.


\acknowledgments

This work supported in part by the U.S Department of Energy under DOE Nuclear Theory Grant DE-FG02-95ER40934.


\end{document}